\documentstyle[12pt,oldlfont]{article}

\begin{document}
\title{PERTURBATIVE QUANTUM FIELD\\ THEORY AT POSITIVE TEMPERATURES:\\
AN AXIOMATIC APPROACH}
\def\theequation{\thesection.\arabic{equation}}

\author{O. Steinmann\\
Universit\"at Bielefeld\\
Fakult\"at f\"ur Physik\\
D-33501 Bielefeld\\
Germany}
\date{Revised version}
\maketitle

\begin{abstract}
It is shown that the perturbative expansions of the correlation functions
of a relativistic quantum field theory at finite temperature are uniquely
determined by the equations of motion and standard axiomatic requirements,
including the KMS condition. An explicit expression as a sum over generalized
Feynman graphs is derived. The canonical formalism is not used, and the
derivation proceeds from the beginning in the thermodynamic limit. No
doubling of fields is invoked. An unsolved problem concerning	existence of
these perturbative expressions is pointed out.
\end{abstract}

\hfill\eject

\section{INTRODUCTION}

The traditional way of describing	thermal equilibrium states of an infinitely
extended quantum system, in particular of a quantum field theory, begins by
restricting the system to a finite volume $V$, defining the canonical or grand
canonical equilibrium by means of the familiar density matrices, and then
going to the limit $V \to \infty$ (the "thermodynamic limit") for the
quantities
for which this limit can be expected to exist [1,2]. This applies especially
to the correlation functions of the fields and closely related objects like
the expectation values of time ordered field products. Up to now most actual
calculations of such functions have been based on this approach, using a
Hamiltonian or Lagrangian formalism at finite $V$.

Another description of equilibria and their local disturbances, which can be
used directly in the thermodynamic limit, has been developed in the framework
of the algebras of local observables [3]. In this approach equilibrium states
are characterized through an analyticity requirement for correlation functions,
the so-called KMS condition. In the present paper we intend to show that this
axiomatic method, suitably adapted to a field theoretical context, is
perfectly capable of handling dynamical problems. More exactly, it will be
shown that perturbative expansions for the correlation functions of a
relativisitic field theory, and related functions, can be derived directly
in the thermodynamic limit, not making use of the canonical formalism, but
using
as only inputs the equations of motion and the axiomatic requirements that
the correlation functions must satisfy. The result is represented as a sum
over generalized Feynman graphs. For the special case of time ordered
functions it agrees with the well-known result of the canonical approach.

Dispensing with the canonical formalism is also a major difference between
our approach and thermo field dynamics [2,4], a Fock space method developed
by H.~Umezawa and coworkers. We differ also from thermo field dynamics by
not invoking a doubling of fields, and by not assigning a basic role to
particles, including quasi-particles. We hold particles to be secondary
objects of the theory, of great phenomenological importance, but little
fundamental significance. In this respect we differ also from the views put
forward by Landsman in ref. [5].

We consider only the $\Phi^4_4$-model, i.e.~the theory of a scalar hermitian
field $\Phi(x)$ satisfying the equation of motion
\begin{equation}
(\Box + m^2) \Phi = - \frac{g}{6} N (\Phi^3) ,
\end{equation}
where $N$ stands for a normal-product prescription taking care of
renormalization. The restriction to $\Phi^4_4$ is merely a matter of
convenience. The generalization of the method to other models, including
gauge theories in local gauges, is straightforward.

We are interested in the correlation functions
\begin{equation}
W(x_1,\ldots, x_n) = \langle\Phi(x_1) \ldots \Phi(x_n) \rangle ,
\end{equation}
where $\langle\cdot\rangle$ denotes the expectation value in a thermal
equilibrium state
with temperature $T\ge 0$. These correlation functions describe the full
physical content of the theory: all observable quantities can in principle
be derived from them (for examples see e.g.~[1,2]). This is so because
knowledge of the $W$ allows the reconstruction of the full representation
of the field algebra by means of the GNS construction [2,3], which yields
a Hilbert space representation of the field algebra with a cyclic vector
$|\rangle$, such that $\langle A\rangle = \langle |R(A)|\rangle$ for any
sufficiently regular function
$A(\Phi)$, with $R(A)$ its representative. All observables of the theory
are supposed to be of this form, and local disturbances of the equilibrium
are created by applying suitable functions $F(\Phi)$ to $|\rangle$.

More generally, we consider the set of functions (or rather, distributions)
\begin{equation}
{\cal W}(X_1, s_1 | \ldots | X_N, s_N) = \langle
T^{s_1} (X_1) \ldots T^{s_N} (X_N)\rangle .
\end{equation}
Here the $X_\alpha$ are non-overlapping sets of 4-vectors $x_i$, the
$s_\alpha$ are signs, and $T^\pm (X)$ denotes respectively
the time-ordered or
anti-time-ordered product of the fields $\Phi (x_i), x_i \in X$.
If each $X_\alpha$ contains only one variable, then $\cal W$ is the correlation
function $W(x_1, \ldots, x_N)$ irrespective of the choice of the signs
$s_\alpha$. For $N=1$ we obtain the usual time-ordered and anti-time-ordered
functions (Green's functions) of the theory. In the sequel the signs
$s_\alpha$ will be frequently suppressed when they are not essential to
understanding.

In a previous work [6], henceforth quoted as $V$, we have derived perturbative
expressions for the functions $\cal W$ in terms of generalized Feynman graphs
for
the case $T=0$, in which case $\langle \cdot \rangle$ denotes the vacuum
expectation value.
This derivation uses neither the Hamiltonian nor the Lagrangian formalism,
but is instead relying on the Wightman axioms as an essential input. We
propose to generalize this method to the case $T>0$. Our way of proceeding
is closely modeled on that taken in V. The ideas and results of that paper
will be freely used. Equation (n.m.) of V will be referred to as eq.
(V.n.m.). The method conists in solving the differential equations for $W$
which
follow from the field equation (1.1), by a power series expansion in the
coupling constant $g$, using the axiomatic properties of $W$ as subsidiary
conditions.

The paper is organized as follows. The assumptions on which the formalism
is based will be stated and discussed in section 2. An explicit formal
expression for the $\cal W$'s as sums over generalized Feynman graphs will be
stated in section 3 and shown to possess the required properties. In section 4
the ultraviolet (UV) divergences of these graphs will be removed by
renormalization. It will be pointed out, however, that renormalization does
not guarantee the existence of the resulting finite-order expressions, on
account of the local singularities of the integrands. This problem remains
unsolved. Finally we will show in section 5 that the expressions of sections
3 and 4, provided they exist, are the only ones satisfying the assumptions
stated in section 2.

\section{ASSUMPTIONS}
\setcounter{equation}{0}

In this section	the assumptions on which our derivations are based,
in addition to the field equation (1.1), will	be enumerated.

The following conditions for $W$ and $\cal W$ are taken over unchanged from
V.

a) The $W$ and $\cal W$ are {\it invariant} under space-time
translations and under space rotations. Invariance under Lorentz boosts
cannot be demanded for $T > 0$..

b) {\it Locality} holds, i.e. $W(X)$ is invariant under the
exchange of two neighbouring variables $x_i, x_{i+1},$ if $x_i - x_{i+1}$
is space-like.

c) The {\it reality condition}
\begin{equation}
{\cal W} (X_1, s_1| \ldots | X_N, s_N)^\ast = {\cal W} (X_N, -s_N|\ldots |
X_1, -s_1)
\end{equation}
holds.

d) The functions $\cal W$ are {\it permutation invariant}
within each sector $X_\alpha$, and they satisfy the {\it splitting
property}
\begin{equation}
{\cal W} (\ldots |X_1 \cup X_2, +|\ldots ) = {\cal W} (\ldots |X_1 ,+|X_2 ,+|
\ldots)
\end{equation}
if $x_i^0 > x_j^0$ for all $x_i \in X_1, x_j \in X_2$. This property shall
hold in every Lorentz frame, not only in the rest frame of the infinite system
under consideration. To that extent we retain Lorentz invariance. These
conditions are not merely a definition of time ordering. That they can be
satisfied is intimately connected with locality.

The {\it cluster property} needs a more careful discussion than it
was accorded in $V$. Let $X, Y,$ be two non-overlapping sets of 4-vectors.
Then the cluster property states that
\begin{equation}
\lim_{|a|\to\infty}	W(X,Y + a) = W(X) W(Y)\ ,
\end{equation}
where $Y + a$ means that all vectors in $Y$ are translated by $a$, and $a$
tends to infinity in a space-like direction. Since derivation with respect
to the coupling constant need not commute with the limit in (2.3), we
cannot expect eq.~(2.3) to hold separately in each order of perturbation
theory, except in the lowest, free, order. But perturbation theory can
also be viewed as an expansion in powers of $\hbar$. And we can demand that
eq.~(2.3) hold for each $W(X)$ in the lowest nonvanishing order in $\hbar$.
This is all that will be needed to establish uniqueness. We set $c=1$, so
that	time and space have the same dimension. The dimension of the field
$\Phi$ is then $\mbox{[mass/length]}^{1/2}$. In order to make the field
equation (1.1) dimensionally consistent, the coupling constant in the
interaction term reads $g/\hbar$, with a dimensionless $g$. We shall
nevertheless set $\hbar = 1$ in our equations, and merely note the
correct $\hbar$-exponents at the points where they are essential.

The spectral condition of the vacuum representation does not hold at
positive temperatures. It is replaced by the {\it KMS condition}, which
we use in its p-space form (see [3], lemma 1.1.1 of chapter V). Define the
Fourier transform $\tilde{\Phi} (p)$ of $\Phi$ by
\begin{equation}
\tilde{\Phi} (p) = (2\pi)^{-5/2} \int dx e^{ipx} \Phi (x),
\end{equation}
and the Fourier transform $\tilde{\cal	W} (P_1 |\ldots |P_N)$ of
${\cal	W} (X_1 |\ldots |X_N)$ accordingly. Let $P_1, \ldots , P_N, Q_{N+1},
\ldots , Q_M , M > N,$ be finite, non-empty sets of 4-momenta. Define
\begin{equation}
P^0 = \sum_{p_i \in \cup P_\alpha} p^0_i .
\end{equation}
Then the KMS condition states that
\begin{equation}
\tilde{\cal W} (P_1 | \ldots |P_N |Q_{N+1} | \ldots | Q_M) = e^{\beta P^0}
\tilde{\cal W} (Q_{N+1}|\ldots |Q_M | P_1 | \ldots | P_N ) ,
\end{equation}
where $\beta = \frac{1}{kT}$ is the inverse temperature.

The normalization conditions stipulated in V are taken over unchanged
for the case $T=0$. They are the standard conditions demanding that $m$ be the
physical mass of the particles associated with the field $\Phi$ at $T=0$,
and defining the coupling constant $g$ and the field normalization in terms
of the Green's functions. In addition we demand that $\Phi$, considered as
an element of an abstract, representation independent, field algebra, is
$T$-independent. This implies that the field equation is $T$-independent,
meaning that the parameters $m$ and $g$ and the subtraction prescription
$N$ do not depend on the temperature.
The prescription $N$ is assumed to be the conventional one. In
particular it should not contain oversubtractions which would destroy
the renormalizability of the theory.
In perturbation theory the
$T$-independence of $N$ means that the UV subtractions and the subsequent
finite renormalizations have the values used at $T=0$ for all $T$. Note that
in the BPHZ method [7], which we will be using,	the subtractions do not
involve prior regularizations or any formal juggling with divergent
quantities. The procedure is therefore well defined. For more details we
refer to section 4. A nonperturbative definition of $N$ can possibly be given
with the help of a Wilson expansion, as explained in section IV.2. of ref.
[7], i.e.~by a point-splitting method with suitable singular coefficient
functions. $T$-independence of $N$ could then be defined as $T$-independence
of these coefficients. We will not explore this possibility any further.

Note that the parameter $m$ is for $T>0$ not the physical mass of a particle.
Quasi-particle masses are determined by the positions of the singularities
of the clothed propagator, which are $T$-dependent.

\section{UNRENORMALIZED SOLUTION}
\setcounter{equation}{0}

The coefficient $W_\sigma (X)$ of order $\sigma$ in the power series
expansion
\begin{equation}
W(X) = \sum^\infty_{\sigma =0}	g^\sigma W_\sigma (X)
\end{equation}
is determined as a solution of the system of differential equations
\begin{equation}
(\Box_i + m^2) \langle \ldots \Phi (x_i) \ldots \rangle_\sigma =
- \frac{1}{6}	\langle\ldots N(\Phi(x_i)^3)\ldots \rangle_{\sigma-1}
\end{equation}
satisfying the subsidiary conditions described in section 2. For $\sigma = 0$
the right-hand side is zero, for $\sigma > 0$ it can be calculated if the
problem has already been solved in order $\sigma -1$.

We first state a formal, unrenormalized, graphical representation for\linebreak
${\cal W}_\sigma (X_1 | \ldots | X_N)$ and then show that it does indeed
satisfy all the requirements.

For the free propagators we use the following notations. We define first
the p-space expressions
\begin{equation}
\tilde{\Delta}_+ (p) = \frac{1}{(2\pi)^3} \delta_+ (p) , \quad
\tilde{\Delta}_F (p) = \frac{i}{(2\pi)^4} \frac{1}{p^2 - m^2+i\epsilon} ,
\end{equation}
which are the usual vacuum propagators, and their thermal extensions
\begin{equation}
\tilde{D}_+ (p) = \tilde{\Delta}_+ (p) + \tilde{C} (p) , \qquad
\tilde{D}_F (p) = \tilde{\Delta}_F (p) + \tilde{C} (p)
\end{equation}
with the $T$-dependent correction term
\begin{equation}
\tilde{C} (p) = \frac{1}{(2\pi)^3} \frac{1}{e^{\beta\omega}-1} \delta
(p^2 - m^2) .
\end{equation}
Here $\delta_+(p) = \theta (p_0) \delta(p^2 - m^2)$ is the $\delta$-function
of the positive mass shell, and $\omega = ({\bf p}^2 + m^2)^{1/2}$. Note that
the
additional term $\tilde{C}$ is the same for $\tilde{D}_+$ and $\tilde{D}_F$.

The $x$-space versions of these propagators are defined by
\begin{equation}
\Delta \ldotp (x) = i \int d^4 p \tilde{\Delta} \ldotp (p) e^{-ipx} , \qquad
D \ldotp	(x) = i \int d^4 p \tilde{D} \ldotp (p) e^{-ipx} .
\end{equation}

The thermal propagators (3.4-5) need not be taken over from the traditional
formalism. They can be derived in our framework by the methods that will
be used in section 5 to establish uniqueness of our solution. But in the
present section we rely on an a posteriori justification, by showing that these
forms give rise to expressions with all the required properties.

${\cal W}_\sigma$ is represented as a sum over generalized Feynman graphs which
are defined as follows. Draw first an ordinary Feynman graph of the $\Phi^4$
theory with $|X| = \sum_\alpha |X_\alpha|$ external	and $\sigma$ internal
vertices. Here $|X_\alpha |$ is the number of points in the set $X_\alpha$.
The graph need not be connected, but must not contain any components without
external points. This graph is called the "scaffolding" of the generalized
graph. Next, it is partitioned into non-overlapping subgraphs, called
"sectors",
such that the external points of a set $X_\alpha$ belong all to the same
sector, but variables in different $X_\alpha$ to different sectors. There may
also exist "internal sectors" not containing external points. The sectors
are either of type $T^+$ or $T^-$. For external sectors this sign is given by
$s_\alpha$. To each sector $S$ we affix its number $\nu (S)$ according to
the following rules.

\begin{itemize}
\item[{i)}]   $\nu (S) = \alpha$ for the external $X_\alpha$-sector.
\item[{ii)}]  If $s_\alpha = s_{\alpha +1}$ there may be an internal sector
with number $\alpha + \frac{1}{2}$. Its type is the reverse of the adjacent
external	sectors: $s_{\alpha + \frac{1}{2}} = - s_\alpha$. If
$s_\alpha	= - s_{\alpha +1}$ no such intermediate internal sector exists.
\item[{iii)}]	If $s_1 = s_N$ there may be an internal sector with number
$N + \frac{1}{2}$ and $s_{N + \frac{1}{2}} = - s_N$. Equivalently we could
give this additional sector the number $\frac{1}{2}$, but we must choose
one of these possibilities and use it consistently.
\end{itemize}

With a partitioned graph we associate a Feynman integrand, which we state
first in $x$-space. To the external points correspond the external variables
$x_i$. To each internal vertex we assign an integration variable $u_j,\quad
j = 1, \ldots, \sigma$. $z_i$ denotes a variable which may be either external
or internal. Within a $T^+$ sector the usual Feynman rules hold: each internal
vertex carries a factor $-ig$, a line connecting the points $z_i$ and $z_j$
carries the propagator $-iD_F (z_i - z_j)$. In a $T^-$ sector the
complex-conjugates of these rules apply. A line connecting points $z_i, z_j$
in different sectors, with $z_i$ lying in the lower-numbered sector, carries
the propagator $-iD_+(z_i - z_j)$. The graph must be divided by the usual
symmetry number if it is invariant under certain permutations of points and
lines.

In p-space, integration variables are assigned to the lines. The vertex
factors in $T^\pm$ factors are $\mp(2\pi)^4g$ and the propagators are
$\tilde{D}_F(p)$ or $(\tilde{D}_F(p))^\ast$ respectively. Lines connecting
different sectors carry propagators $\tilde{D}_+(p)$, the momentum $p$
flowing from the lower to the higher sector. For each external point there
is a factor $(2\pi)^{3/2}$. Momentum is conserved at each internal vertex.

For $T=0$ the rule ii) governing internal sectors seems to differ from the
corresponding rule in V, where instead of one (possibly empty) intermediate
sector of different type we had chains of intermediate sectors of the same
type as the bracketing external sectors. However, it can be shown with the
help of Lemma V.3.1 that the two formulations are equivalent. Graphs
containing non-empty internal sectors according to iii) vanish for $T=0$,
wherefore they do not occur in V.

For the Green's functions $(N=1)$ our rules agree with those of the
conventional
real-time formalism. Our graphs are equal to those of the Keldysh formulation
[8].

\bigskip
The first point to be checked is that the above prescription gives an
unambiguous result for the correlation functions $W$. The problem is that in
the definition $W(x_1, \ldots , x_n) = {\cal W}(x_1, s_1 | \ldots | x_n , s_n
)$ the
signs $s_\alpha$ can be chosen arbitrarily. But it can be shown that the
sum over all graphs with the same scaffolding does not depend on the choice
of these signs. The proof is modeled closely on the corresponding proof in
V and will only be briefly indicated. Consider a given scaffolding.
Let $S$ denote an internal $T^+$ sector, $S_x$ an external $T^+$ sector with
external variable $x, \bar{S}$ and $\bar{S}_x$\,  $T^-$ sectors. A product
$S_xS$ or the like denotes the sum over all partitions of a given subgraph
into two adjacent sectors $S_x, S$. The propagators connecting these two
sectors are included. We will first show that the choice of $s_N$ is
irrelevant.
The graphs belonging to $s_N = +$ or $s_N = -$ respectively differ only in the
sectors with number $\nu (S) > N-1$. In the case $s_{N-1} = +, s_1 = -$
these variable sectors are (we put $x_N = x$) $S_x + \bar{SS}_x$ for
$s_N = +$, and $\bar{S}_x + \bar{S}_x S$ for $s_N = -$. But
\[
S_x + \bar{SS}_x = \bar{S}_x + \bar{S}_x S
\]
is a special case of Lemma V.3.1, which remains valid for our new propagators.
Similarly, if $s_1 = s_{N-1} = +$ we must show that
\[
S_x + S_x \bar{S} + \bar{SS}_x + \bar{SS}_x \bar{S} = \bar{S}_x	,
\]
or
\[
(S_x + S_x \bar{S} - \bar{S}_x - \bar{SS}_x) + \bar{S} (S_x + S_x\bar{S} -
\bar{S}_x - \bar{SS}_x) + (S+\bar{S}+\bar{SS}) \bar{S}_x = 0 ,
\]
which is correct because all three brackets vanish as a result of Lemma
V.3.1. The proof of independence on the other $s_\alpha$ follows similar
lines. The case $1 < \alpha < N$ is considerably simplified compared to V
by the new formulation of the rules concerning internal sectors.

Of the conditions stated in section 2 invariance under translations and
rotations, and the symmetry of $\cal W$ within $T^\pm$ factors are trivially
satisfied. That the equation of motion (3.2) is satisfied, is shown
exactly as in V. At the present formal level the product $N(\Phi^3)$ denotes
Wick ordering $:\Phi^3(x):$ with respect to the free vacuum field.

The proof of the reality condition (2.1) and Ostendorf's proof [9] of the
splitting relation (2.2) can also been taken over unchanged from V. Since the
non-invariant part $C(x-y)$ is the same in $D_+$ and in $D_F$, the crucial
relation $D_F(x-y) = D_+ (x-y)$ if $x^0 > y^0$ holds in every orthochronous
Lorentz frame. Hence the same is true for the splitting relation. From this
and the symmetry within sectors we can prove locality. Let $(x-y)^2 < 0$.
Then there exist two Lorentz frames such that in one $x^0 > y^0$, in the
other $x^0 < y^0$. Hence
\begin{eqnarray*}
& W_\sigma(\ldots, x, y, \ldots ) =	{\cal W}_\sigma (\ldots | x,y,+|
\ldots ) = \nonumber \\
&= {\cal W}_\sigma (\ldots |y, x, + | \ldots ) = W_\sigma
(\ldots , y, x, \ldots ).
\end{eqnarray*}

Our propagators, being the 2-point functions of a free field, contain
for dimensional reasons a factor $\hbar$ (this refers to x-space). The
vertices carry a factor $\hbar^{-1}$, so that the $2n$-point function
of order $\sigma$ contains a factor $\hbar^{n+\sigma}$. The term
of lowest order in $\hbar$ is therefore the free $\sigma = 0$
expression. It is simply a sum over all possible products of free
2-point functions, and obviously satisfies the cluster property, because
$D_+ (\xi)$ and $D_F(\xi)$ converge to zero if $\xi$ tends to infinity
in a space-like direction. Hence our weak version of the cluster property
is satisfied.

Yet to be proved remains the KMS condition (2.6). It is easy to see that a
graph contributing to the left-hand side of (2.6) becomes a graph
contributing to the right-hand side, if the directions of all lines connecting
sectors with $\nu (S) < N+1$ to sectors with $\nu (S) \ge N+1$ are reversed,
and vice versa. This means that the corresponding propagators $\tilde{D}_+ (k)$
are replaced by $\tilde{D}_+ (-k)$. From the definitions (3.3-5) we find
$\tilde{D}_+ (k) = e^{\beta k_0} \tilde{D}_+ (-k)$. Hence the two variants
of the considered graph differ by the factor $\exp (\beta \sum_\alpha
k^\alpha_0)$, the sum extending over the momenta of the lines in question.
But $K = \sum k^\alpha$ is the total momentum flowing from the $P$-part of
the graph to the $Q$-part, so that $K^0 = P^0$ as defined in (2.5). As a
side remark we note that a similar argument can be used to show that it is
immaterial whether extremal internal sectors are assigned	the number
$\frac{1}{2}$ or $N + \frac{1}{2}$.

\section{RENORMALIZATION AND THE
EXIS\-TENCE PROBLEM}
\setcounter{equation}{0}

As yet, the expressions derived in the preceding section have only a formal
meaning. There remains the question of the existence of the integrals
symbolized by the graphs.

The UV	problem is concerned with the behaviour of the integrands at infinity
in momentum space. It can be handled exactly as was done in V for the vacuum
case. We note first that loop integrals over loops extending over more than
one sector are finite because of the strong decrease of $\tilde{D}_+ (k)$
for $k_0 \to - \infty$ and momentum conservation: primitively divergent
subgraphs exist only within sectors, where they can be treated with
conventional methods. We choose BPHZ renormalization (see [7]), which
introduces suitable subtractions in the integrand, before integrating,
thereby avoiding the need for regularization. The subtractions are found by
expanding the integrands of potentially dangerous subgraphs, the
"renormalization parts", into power series of sufficiently high degree in their
external momenta. The temperature dependent part $\tilde{C}$ of the
propagator $\tilde{D}_F$ decreases exponentially fast at large momenta and is
innocent of any UV problems. We can therefore define the mentioned subtraction
terms using $\tilde{\Delta}_F$ instead of $\tilde{D}_F$ inside the
renormalization parts, without destroying the UV convergence achieved by the
subtraction. We also define the finite renormalizations of the BPHZ
prescription
to be $T$-independent, giving them the values needed to satisfy the
normalization
conditions at $T=0$ stated in V. These rules are the expression of the
required $T$-independence of the renormalization prescription $N$ in the
field equation (1.1). An important effect of this limited subtraction is the
emergence for $T>0$, of lines connecting a vertex to itself. But they carry
the propagator $\tilde{C} (k)$, hence the loop integral over $k$ exists,
and is the same in $T^+$ and in $T^-$ sectors.

But the UV problem is not the only existence problem we are faced with.
There is also the problem of the local singularities of the integrand,
again in $p$-space. The integrand is a product of distributions in variables
which are quadratic functions of a complete set of independent external
and internal momenta. The variables in a given subset of propagators may
be dependent (i.e.~their gradients in momentum space may be linearly
dependent) on certain manifolds, in which case the product of propagators
does not necessarily define a distribution. It is no longer clearly integrable
even in the sense of distributions. Contrary to assertions found in the
literature, the problem is not restricted to the case of two propagators
separated by a self energy insertion, and thus both depending on the same
variable. As an example of a more complex situation, consider the integral
$\int dk \tilde{D}_\cdot (k) \tilde{D} \cdot (p-k)$ over a two-line loop,
the dots
standing for either $+$ or $F$. The mass shell $\delta 's$ in the two factors
coalesce at $p=0$, hence the integral diverges at that point, and this
singularity in the external variable $p$ (external to the considered 2-line
subgraph) is not removed by renormalization. Closer inspection shows that the
singularity is of first order. A chain of $n$ two-line bubbles will then
produce a singularity of order $|p|^{-n}$, which is not integrable for $n\ge
4$.
Nor is it defined in another way as a distribution. The remaining integration
over $p$, which may be an internal or external variable of the full graph,
is therefore not defined: individual graphs containing such chains diverge.
These divergences may cancel between graphs with the same scaffolding but
different sector assignments of the vertices in the chain. But we are not
aware of a proof to this effect, even for this still relatively simple
example.

How is this problem solved in the vacuum case? This is easiest for the fully
time ordered function $\tilde{\tau} (P) = \tilde{\cal W} (P, +)$. If $m > 0$
this function is everywhere the boundary value of an analytic function in
complexified variables $p_i$. The same is true for the Feynman propagator
$(k^2 - m^2 + i \epsilon )^{-1}$. For variables $\{ p_i \}$ in the domain of
analyticity of $\tilde{\tau}$ we can deform the integration contours for the
internal variables $k_j$ into the complex in such a way that they never meet
the singularities at $k_j^2 = m^2$. The integrand is then a smooth function,
and there are	no problems with the local existence of the integral. But the
presence of a $\delta$-term in $\tilde{D}_F$ destroys analyticity, so that
the method does not work for positive $T$. For $m=0,\quad \tilde{\Delta}_F (k)$
is singular at $k=0$ and is there not a boundary value of an analytic function.
The same holds for $\tilde{\tau}$ at points where a partial sum of $p_i$'s
vanishes. These singularities can lead to infrared divergences, which need
special attention. An $x$-space method for achieving this, and also for
proving existence of the general ${\cal W}_\sigma$, has been described in V.
It relies heavily on the fact that $\Delta_+ (\xi )$ is analytic in $Im \xi^0
< 0$ and decreases at least of second order for $Im \xi^0 \to - \infty$.
But $D_+ (\xi )$ is only analytic in the strip $-\beta < Im \xi^0 < 0$, so
that this method can also not be extended to positive $T$.

At present the problem of existence of ${\cal W}_\sigma$ remains unsolved.

\section{UNIQUENESS}
\setcounter{equation}{0}

It will be shown that the expansion described in sections 3 and 4, assuming
its existence in finite orders, is the only expression satisfying all the
requirements.

Assume that uniqueness has been established up to order $\sigma - 1$. Let
$W_\sigma^1, W_\sigma^2 ,$ be two solutions of the equations (3.2) plus
subsidiary conditions. Then their difference
\begin{equation}
h_\sigma (x_1, \ldots , x_n) = W^{1}_\sigma (x_1, \ldots) - W^{2}_\sigma
(x_1, \ldots)
\end{equation}
satisfies the homogeneous equations
\begin{equation}
(\Box_i + m^2) h_\sigma (\ldots , x_i , \ldots ) = 0
\end{equation}
and all the subsidiary conditions.
We want to derive the most general form of $h_\sigma$.

The Fourier transform ${\tilde h}_\sigma (p_1, \ldots, p_n)$ must contain
a factor $(p_i^2 - m^2)$ for each $i$.

Define
\begin{equation}
h_\sigma ( \ldots [x,y] \ldots ) = h_\sigma (\ldots, x, y \ldots	) - h_\sigma
(\ldots y, x \ldots)
\end{equation}
and let ${\tilde h}_\sigma (\ldots [p,q] \ldots )$ be its Fourier transform.
As in $V$ we can prove that ${\tilde h}_\sigma (\ldots [p,q] \ldots)$ has
its support contained in the manifold $p+q=0$, and must therefore be of the
form
\begin{equation}
{\hat h}_\sigma (\ldots [p,q] \ldots) = \delta^4 (p+q) \delta (p^2 - m^2)
\{ f_+	({\bf p}) + \epsilon (p_0) f_- ({\bf p}) \}			\ ,
\end{equation}
where the dependence on the variables indicated by dots has been suppressed.
By locality the support of the function $\theta (x^0 - y^0) h_\sigma (\ldots
[x,y] \ldots)$ is contained in the set $(x-y) \epsilon \overline{V}_+$,
the closed forward cone. Its Fourier transform
\[
\frac{i}{2\pi} \delta^4 (p+q) \left[ \frac{f_+ ({\bf p})}{\omega ({\bf p})}
\frac{p_0}{(p_0 + i\epsilon)^2 - {\bf p}^2 - m^2} + f_- ({\bf p})
\frac{1}{(p_0 + i\epsilon)^2 - {\bf p}^2 - m^2} \right] \ ,
\]
considered as a function of $p+q$ and $q$, is in $p$ analytic in the forward
tube $\{ \mbox{Im} p \in V_+\}$ and is polynomially bounded in the slightly
smaller tube $\{ (  \mbox{Im} p - a) \in V_+\}$ for any $a \in V_+$ [10]. The
$p_0$-dependent	factors in the above expression have the correct analyticity.
In order for this to be the case for the full expression, the ${\bf
p}$-dependent terms
$\omega^{-1} f_+$ and $f_-$ must be entire functions of ${\bf p}$.
Furthermore, they must be polynomially bounded, hence they are polynomials.
As a result we find
\begin{equation}
h_\sigma (\ldots [p,q] \ldots) = \delta^4 (p+q) \delta (p^2 - m^2)
\{ \omega ({\bf p}) F ({\bf p}) + \epsilon (p_0) G({\bf p}) \}
\end{equation}
with $F$ and $G$ polynomials in ${\bf p}$. $F$ and $G$ may also depend on the
variables not shown explicitly, and factors only depending on those other
variables have been omitted.

Consider next the double-commutator function
\begin{equation}
h_\sigma (\ldots [[x,y], z] \ldots ) = h_\sigma (\ldots [x,y] z \ldots)
- h_\sigma (\ldots z[x,y] \ldots ) \ .
\end{equation}
By locality	its support is contained in the set $S = \{ (x-z)^2 \ge 0\
\mbox{or}\ (y-z)^2 \ge 0 \}$. But by transforming (5.5) into $x$-space
we find that the right-hand side of (5.6) depends on $x$ and $y$ only in the
combination $x-y$, i.e.~it is invariant under simultaneous translation
of $x$ and $y$ by the same 4-vector.	But no subset of $S$, apart from the
empty set, is invariant under such a simultaneous translation. Therefore
we find
\begin{equation}
h_\sigma (\ldots [[x,y]z] \ldots ) = 0.
\end{equation}

Finally we find from the KMS condition (2.6) that
\begin{eqnarray}
(1-\exp (-\beta p^0_1)) {\tilde h}_\sigma (p_1, \ldots, p_n) & = &
{\tilde h}_\sigma (p_1, \ldots, p_n) - {\tilde h}_\sigma (p_2, \ldots,
p_n, p_1) \nonumber\\
& = & \sum^n_{i=2} {\tilde h}_\sigma (p_2, \ldots [p_1, p_i] \ldots, p_n),
\end{eqnarray}
from which ${\tilde h}_\sigma$ can be obtained through division by
$(1-\exp (-\beta p^0_1))$. Reinserting the resulting expression in the terms
on the right-hand side of eq.~(5.8), and iterating this procedure
sufficiently many times, we find that ${\tilde h}_\sigma (p_1, \ldots,
p_{2n})$ must be of the form
\begin{eqnarray}
& {\tilde h}_\sigma (p_1, \ldots, p_{2n}) \nonumber\\
&=& \sum F ({\bf p}_{i_1}, \ldots, {\bf p}_{i_n}) \prod^n_{\alpha=1}
[ 1 - \exp ( - \beta p^0_{i_\alpha} ) ]^{-1} \delta (p^2_{i_\alpha} - m^2)
\delta^4 (p_{i_\alpha}	+ p_{j_\alpha})
\end{eqnarray}
where the sum extends over all partitions of the $2n$ variables into $n$
ordered pairs $(p_{i_\alpha} , p_{j_\alpha}	),\ i_\alpha < j_\alpha,$
and $F$ is a polynomial in ${\bf p}_i$ and $\omega ({\bf p}_i)$. The
odd-point functions vanish in $\Phi^4$ theories.

Now, if $F$ is a genuine polynomial, not a constant, then the expression
(5.9) shows a bad high-energy behaviour for $p^0_{i_\alpha} \to \infty$,
which would destroy renormalizability. Hence $F$ must be a constant, and
$h_\sigma$ has in x-space the form
\begin{equation}
h_\sigma (x_1, \ldots, x_{2n} ) = c_{\sigma n} \sum_{\mbox{pairings}}
\prod_\alpha	D_+ (x_{i_\alpha}	- x_{j_\alpha}) .
\end{equation}
For dimensional reasons $c_{\sigma n}$ must contain a factor $\hbar^n$,
independently of $\sigma$. Insertion of such expressions in the right-hand
side of the equation of motion (3.2) leads to terms of higher order in
$\hbar$. This means that the homogeneous terms (5.10) are the contributions
of lowest order to the $2n$-point functions. Since their functional form
does not depend on $\sigma$, their contributions to $W(x_1, \ldots, x_{2n})$
can be summed over $\sigma$, in the sense of formal power series, to yield
\begin{equation}
c_n (g) \sum \prod_\alpha D_+ (\ldots)
\end{equation}
as term of lowest order in $\hbar$. The normalization condition for the
$2$-point	function determines the value of $c_1$ for $T=0$ to be $c_1 = -i$.
With a different choice of $c_1$ for $T>0$ our $T$-independent
definition of the normal product $N$ would no longer remove the ultraviolet
divergences, i.e.~such a choice would destroy the validity of the field
equation (1.1). Hence we must have
\begin{equation}
c_1 = - i
\end{equation}
identically in $g$ and $T$.

By our assumptions the expression (5.11) must satisfy the cluster property
(2.3). Using induction with respect to $n$ one finds the unambiguous
result
\begin{equation}
c_n = (-i)^n\ .
\end{equation}
This value does not depend on $g$, which proves the desired unicity
\begin{equation}
h_\sigma = 0
\end{equation}
for $\sigma > 0$. In the lowest order $\sigma = 0, h_0$ is the only
nonvanishing contribution to the $2n$-point function. It has the conventional
free-field form, as has already been anticipated in section 3.

It is a curious aspect of this argumentation that it does not work in the free
case $g=0$, where renormalizability is not at stake. For the free theory one
could
arrive at the same results faster by postulating the usual canonical
commutation
relations. For interacting fields this method is rather less convincing,
because
in that case the CCR's have no rigorous meaning, since interacting fields
cannot be restricted to sharp times.

\bigskip
\noindent
\section{ACKNOWLEDGEMENTS}

I am indebted to R.~Baier and A.K.~Rebhan for enlightening discussions
and comments, and for guidance to the literature on finite-temperature
field theory.

\bigskip
\noindent
\section{REFERENCES}
\noindent
\begin{enumerate}
\item J.I.~Kapusta: Finite-Temperature Field Theory. Cambridge University
      Press, Cambridge, 1989.

\item N.P.~Landsman and Ch.G.~van Weert: Real- and imaginary-time field
theory at finite temperature and density.
Phys.~Reps.~145, 141 (1987).

\item R.~Haag: Local Quantum Physics. Springer, Berlin 1992.

\item H.~Umezawa, H.~Matsumoto, and M.~Tachiki: Thermo Field Dynamics and
      Condensed States. North-Holland, Amsterdam 1982.\\
      H.~Umezawa: Advanced Field Theory. American Institute of Physics,
      New York 1993.

\item N.P.~Landsman: Non-shell unstable particles in thermal field theory.
Ann.~Phys. (NY) 186, 141 (1988).

\item O.~Steinmann: Perturbation theory of Wightman functions.
Commun.~Math.~Phys.~152, 627 (1993).

\item W.~Zimmermann, in: Local operator products and renormalization in
quantum field theory.
Lectures on Elementary Particles and Quantum Field
      Theory. Eds.~S.~Deser and H.~Pendleton. MIT Press, Cambridge MA 1971.

\item L.V.~Keldysh: Diagram technique for nonequilibrium processes.
Sov. Phys.~JETP 20, 1018 (1964).

\item A.~Ostendorf: Feynman rules for Wightman functions.
Ann.~Inst.~H.~Poincar\'e 40, 273 (1984).

\item R.F.~Streater and A.S.~Wightman: PCT, Spin \& Statistics, and All That,
      chapter 2-3. Benjamin/Cummings, Reading MA 1978.

\end{enumerate}

\end{document}